\input{aipcheck.tex}
\documentclass[final]{aipproc}
\layoutstyle{6x9}
\def\be{\begin{equation}}
\def\ee{\end{equation}}
\def\bea{\begin{eqnarray}}
\def\eea{\end{eqnarray}}
\def\ba{\begin{array}}
\def\ea{\end{array}}

\begin{document}
\title[Meson-nucleon sigma terms]{Strangeness and meson-nucleon sigma terms}
\author{Harleen Dahiya}{address={Department of Physics, Dr. B.R. Ambedkar National Institute of Technology,
Jalandhar, Punjab-144 011, India.}}
\author{Neetika Sharma}{address={Department of Physics, Dr. B.R. Ambedkar National Institute of Technology,
Jalandhar, Punjab-144 011, India.}}

\date{\today}

\begin{abstract}

The chiral constituent quark model ($\chi$CQM)  has been extended
to  calculate the flavor structure of the nucleon through the
meson-nucleon sigma terms which have large contributions from the
quark sea and are greatly affected by  chiral symmetry breaking
and SU(3) symmetry breaking. The hidden strangeness component in
the nucleon has also been investigated and its significant
contribution is found to be consistent with the recent available
experimental observations.

\end{abstract}

\maketitle

The knowledge on the internal structure of the nucleon has been
rather limited because of confinement and it is still a big
challenge to perform the calculations from the first principles of
Quantum Chromodynamics (QCD).  The measurements of polarized
structure functions of proton in the deep inelastic scattering
(DIS) experiments \cite{emc} provided the first evidence that the
valence quarks of proton carry only a small fraction of its spin
in contradiction with the predictions of the Naive Quark Model
(NQM) \cite{nqm}.

Several interesting facts have also been revealed regarding the
flavor distribution functions \cite{nmc} indicating that the
structure of the nucleon is not limited to $u$ and $d$ quarks
only. Recently, there have been indications of non-zero
strangeness content in the nucleon by the experiments measuring
electromagnetic form factors \cite{sample} as well as in the
context of low-energy experiments \cite{nutev,pion nucleon}.

The meson-nucleon sigma terms \cite{pion nucleon} are the
fundamental parameters to test the chiral symmetry breaking
effects and thereby determine the scalar quark content of the
baryons.  They are theoretically interesting because they are
known to have intimate connection with the dynamics of the
non-valence quarks at low-energy \cite{gasser,riaz,bass}.

We plan to understand the implications of chiral symmetry breaking
for the scalar matrix elements of the nucleon within the chiral
constituent quark model ($\chi$CQM)
\cite{manohar,hd,hdmagnetic,hdasymmetry}. In particular, we would
like to phenomenologically estimate the quantities affected by the
hidden strangeness component in the nucleon as well as to study
the meson-nucleon sigma terms and the meson-baryon sigma terms for
$\Sigma$ and $\Xi$ baryons which are expected to have large
contributions from the quark sea.

The key to understand the $\chi$CQM formalism \cite{cheng}, is the
fluctuation process $ q^{\pm} \rightarrow {\rm GB} + q^{' \mp}
\rightarrow (q \bar q^{'}) +q^{'\mp}$, where $q \bar q^{'} +q^{'}$
constitute the ``quark sea'' \cite{hd,cheng,johan}. The effective
Lagrangian describing the interaction between quarks and a nonet
of GBs  can be expressed as ${\cal L}= g_8 {\bf \bar
q}\left(\Phi+\zeta\frac{\eta'}{\sqrt 3}I \right) {\bf q}=g_8 {\bf
\bar q}\left(\Phi' \right) {\bf q}$.

The flavor structure of the nucleon is defined as \cite{cheng}
$\hat N \equiv \langle N |q \bar q| N \rangle$ where $|N\rangle$
is the nucleon wavefunction and $q {\bar q}$ is the number
operator for the scalar quark content measuring the sum of the
quark and antiquark numbers. The pion-nucleon sigma term
($\sigma_{\pi N}$) affected by the contributions of the quark sea
is expressed as  \bea \sigma_{\pi N} = \hat m \frac{ \langle N|
{\bar{u}u} + {\bar{d}d} -2 {\bar{s}s} | N \rangle}{1- 2y_N}
=\frac{\hat {\sigma}}{1-2 y_N} \label{fs1} \,, \eea where we have
defined \be \hat {\sigma} = {\hat m}{\langle N| {\bar{u}u} +
{\bar{d}d} -2 {\bar{s}s} | N \rangle} ~~~~~{\rm and} ~~~~~y_N =
\frac{\langle N|{\bar{s}} s |N \rangle} {\langle N| {\bar{u}u} +
{\bar{d}d} | N \rangle}\,. \ee

The strangeness fraction of the nucleon and strangeness sigma term
are respectively defined as \be f_s = \frac{\langle N|{\bar{s}} s
|N \rangle} {\langle N| {\bar{u}u} + {\bar{d}d} + {\bar{s}s} | N
\rangle} = \frac{\sigma_{\pi N} - \hat{\sigma}}{3 \sigma_{\pi N} -
\hat{\sigma}} ~~~~~{\rm and} ~~~~~ \sigma_s=m_s {\langle N|
\bar{s}s | N \rangle} =\frac{1}{2}y_N \frac{m_s}{\hat m}
\sigma_{\pi N}\,. \ee Further, the sigma terms corresponding to
the strange mesons can be expressed as \bea \sigma_{K N} &=&
\frac{\sigma^u_{K N}+\sigma^d_{K N}}{2} = \frac{\hat{m} +
m_s}{2}\langle N| {\bar{u}u} + {\bar{d}d}+2 {\bar{s}s}|N\rangle
=\frac{\hat m +m_s}{4 \hat m}(2 \sigma_{\pi N} - \hat{\sigma}) \,,
\eea  \bea \sigma_{\eta N} &=& \frac{1}{3} \langle
N|\hat{m}({\bar{u}u}+{\bar{d}d})+ 2 m_s
{\bar{s}s}|N\rangle  
= \frac{1}{3} \hat{\sigma}+ \frac{2 (m_s+ {\hat m})}{3 {\hat m
}}y_N \sigma_{\pi N} \,. \eea

The SU(3) symmetric and antisymmetric scalar matrix elements
characterizing the weak matrix elements for the flavor structure
are expressed as \be F_S = \frac{1}{2} \langle N | {\bar u} u -
{\bar s} s|N \rangle,~~~~~~~~ D_S = \frac{1}{2} \langle N | {\bar
u} u -2 {\bar d} d + {\bar s} s|N \rangle\,.\ee Similarly, the
singlet and non-singlet combinations of the  flavor structure can
be related to the weak couplings and are expressed as \be g^0_A =
\langle N | {\bar u} u + {\bar d} d + {\bar s} s |N
\rangle,~~~~~g^3_A = \langle N | {\bar u} u - {\bar d} d|N
\rangle,~~~~~ g^8_A =  \langle N | {\bar u} u + {\bar d} d -2
{\bar s} s |N \rangle. \ee

In Table \ref{strange}, we have presented the results of our
calculations pertaining to the scalar matrix elements affected by
the strangeness content of the nucleon. To understand the
implications of the strange quark mass and SU(3) symmetry
breaking, we have presented the results with and without SU(3)
symmetry breaking. From the Table one finds that the present
result for the strangeness content in the nucleon $y_N$ and
strangeness fraction of the nucleon $f_s$ looks to be in agreement
with the most recent experimental and phenomenological results
available. The non-zero values for $y_N$ and $f_s$ in the present
case indicate that the chiral symmetry breaking is essential to
understand the significant role played by the quark sea. It is
also clear from the table that, in general, the quantities
involving the strange quark content are very sensitive to SU(3)
symmetry breaking. For example, the values of the strangeness
dependent quantities $y_N$, $f_s$, $F_s$, $D_s$, $G_s^0$, $G_s^8$,
$\hat {\sigma}$, $\sigma_s$, $\sigma_{\pi N}$, $\sigma_{K N}$, and
$\sigma_{\eta N}$ change to a large extent when compared for the
SU(3) symmetric and SU(3) symmetry breaking case. The results for
other quantities which do not have strangeness contribution are
not much different for both the cases.

For $\sigma_{\pi N}$, the value of $\chi$CQM with SU(3) symmetry
can give a value in the higher range by adopting a larger value of
$\hat {\sigma}$ however, as has been shown in our earlier work
\cite{hd}, SU(3) symmetry does not give a satisfactory description
of quark sea asymmetry and spin related quantities.  A refinement
in the analysis of $\pi-N$ scattering giving higher values of
$\sigma_{\pi N}$ would not only strengthen the mechanism of chiral
symmetry breaking generating the appropriate amount of strangeness
in the nucleon but would also justify the consequences of SU(3)
symmetry breaking mechanism.

The calculations can be extended to predict the meson-nucleon
sigma terms ($\sigma_{KN}$ and $\sigma_{\eta N}$) as well as the
meson-baryon sigma terms for $\Sigma$ and $\Xi$ baryons (Table
\ref{baryon}). The $\sigma_{K N}$ and $\sigma_{\eta N}$ terms are
found to be quite sensitive to $y_N$ and also become strangely
large for the SU(3) symmetric case. The future DA$\Phi$NE
experiments \cite{dafne} to determine KN sigma terms could
restrict the model parameters and provide better knowledge of
strangeness content of the nucleon which would also have important
implications for the hyperon-antihyperon production in the heavy
ion collisions.

In conclusion, we would like to state that chiral symmetry
breaking is the key to understand the hidden strangeness content
of the nucleon. In the nonperturbative regime of QCD, constituent
quarks and the weakly interacting Goldstone bosons constitute the
appropriate degrees of freedom at the leading order.

\section{ACKNOWLEDGMENTS}
H.D. would like to thank Department of Science and Technology,
Government of India (SR/S2/HEP-0028/2008) and the organizers of
BARYON'S10 for financial support.

\begin{table}
\begin{tabular}{|c|c|c|c|c|} \hline
 &  & NQM & \multicolumn{2}{c|}{$\chi$CQM}
\\ \cline{4-5}
Quantity& Phenomenology & \cite{nqm} & with SU(3)&{ with SU(3)}
\\& & & symmetry &{symmetry breaking} \\ \hline $\langle N |{\bar
u} u| N \rangle$ &...&$ \leq $ 2&2.41&2.44 \\ $\langle N |{\bar d}
d| N \rangle$ &...&$ \leq $ 1&1.75&1.68\\ $\langle N |{\bar s} s|
N \rangle$ &...& 0.0&1.08&0.18\\
$y_N$ & $0.11 \pm 0.07$ \cite{gasser}& 0.0 & 0.26 &0.044 \\ $f_s$
&$0.10 \pm 0.06$ \cite{nutev}& 0.0 & 0.21 & 0.042
\\ $F_s$ & 1.52 \cite{maiani}, ~ 1.81 \cite{gasser}& $ \leq $ 1
&0.67& 1.13 \\ $D_s $ & $-0.52$ \cite{maiani}, ~ $-$0.57
\cite{gasser}& 0.0 & 0.0 & $-$0.37
\\ $R_s$ &...& $ \leq $ 3 & 6.22 & 5.39 \\ $G_S^0$ &...& $ \leq $ 3 & 5.24
& 4.30 \\ $G_S^3$ & ...& $ \leq $ 1 & 0.67 & 0.76 \\ $G_S^8$
&...&$ \leq $ 3&2.01 &3.76 \\ ${\hat \sigma}$ &...&
28.57&28.57&28.57 \\ $\sigma_{s}$ &...&0 &168.71&15.12 \\
$\sigma_{\pi N}$ &...&28.57&59.25&31.32\\ $\sigma_{K N}$
&...&164.29&517.04&195.90\\
$\sigma_{\eta N}$  &...&9.52& 244.70&30.60 \\ \hline
\end{tabular}
\caption{ The $\chi$CQM results for the scalar matrix elements of
the nucleon.} \label{strange}
\end{table}

\begin{table}
\begin{tabular}{|c|c|c|c|} \hline
Baryon (B) & $\sigma_{\pi B}$ & $\sigma_{K B}$ & $\sigma_{\eta B}$
\\ \hline $N$&  31.32 & 195.90 & 30.60 \\ $\Sigma$&
137.76& 1419.97&846.65\\ $\Xi$&$-$17.96 & $-$370.76&$-$347.17 \\
\hline
\end{tabular}
\caption{The $\chi$CQM results for the meson-baryon sigma terms.}
\label{baryon}
\end{table}

\end{document}